\documentclass[journal]{IEEEtran}

\usepackage{graphicx} 
\usepackage{cite} 
\usepackage{amsmath,amssymb} 
\usepackage{breqn} 
\usepackage{booktabs} 
\usepackage{tablefootnote} 
\usepackage{tikz} 
\usetikzlibrary{positioning, calc, arrows.meta, shapes.geometric, decorations}
\usepackage[breaklinks]{hyperref} 
\usepackage{url} 

\graphicspath{{../Figures/}}
\DeclareGraphicsExtensions{.pdf,.eps}

\hypersetup{
  breaklinks=true,
  colorlinks=false,
  urlcolor=black
}
    \tikzset{
        block/.style = {draw, rectangle,
            minimum height=1cm,
            minimum width=2cm},
        input/.style = {coordinate,node distance=1cm},
        output/.style = {coordinate,node distance=4cm},
        arrow/.style={draw, -latex,node distance=2cm},
        pinstyle/.style = {pin edge={latex-, black,node distance=2cm}},
        sum/.style = {draw, circle, node distance=1cm},
    }
\tikzstyle{process} = [rectangle, minimum width=3cm, minimum height=1cm, text centered, draw=black, fill=blue!10]
\tikzstyle{startstop} = [ellipse, minimum width=3cm, minimum height=1cm, text centered, draw=black, fill=green!20]
\tikzstyle{arrow} = [thick,->,>=Stealth]

\begin{document}
\title{A Mission Engineering Framework for Uncrewed Aerial Vehicle Design in GNSS-Denied Environments for Intelligence, Surveillance, and Reconnaissance Mission Sets}

\author{Alfonso Sciacchitano, 
        Douglas L. Van Bossuyt
\thanks{Manuscript received Month DD, YYYY; revised Month DD, YYYY; accepted Month DD, YYYY. (Corresponding author: Douglas L. Van Bossuyt.)}
\thanks{A. Sciacchitano is with the Department of Mechanical and Aerospace Engineering at the Naval Postgraduate School (e-mail: 
alfonso.sciacchitano@nps.edu).}
\thanks{D.L. Van Bossuyt is with the Department of Systems Engineering at the Naval Postgraduate School (e-mail: douglas.vanbossuyt@nps.edu).}
}


\maketitle

\begin{abstract}
Small, low–size, weight, power, and cost (SWaP-C) uncrewed aerial vehicles (UAVs) are increasingly used for intelligence, surveillance, and reconnaissance (ISR) missions due to their affordability, attritability, and suitability for distributed operations. However, their design poses challenges including limited endurance, constrained payload capacity, and reliance on simple sensing modalities such as fixed–field-of-view, bearing-only cameras. Traditional platform-centric methods cannot capture the coupled performance, cost, and coordination trade-offs that emerge at the system-of-systems level.

This paper presents a mission engineering framework for early-phase design of low-SWaP-C UAV ISR architectures. The framework integrates design of experiments, multi-objective optimization, and high-fidelity simulation into a closed-loop process linking design variables to estimator-informed performance and mission cost. Candidate architectures are explored via Latin hypercube sampling and refined using a genetic algorithm, with performance evaluated through Monte Carlo trials of a federated Kalman filter benchmarked against the posterior Cram\'er–Rao lower bound. Validation follows the Validation Square methodology, combining theoretical, empirical, and structural assessments.

A case study on man-overboard localization in a GNSS-denied maritime environment shows that localization accuracy saturates at sub-meter levels, while higher-cost configurations primarily add redundancy and resilience. The framework thus quantifies mission trade-offs between performance, affordability, and robustness, providing a scalable decision-support tool for contested, resource-constrained ISR missions.
\end{abstract}




\section{Introduction}
Small uncrewed aerial vehicles (UAVs) with low size, weight, power, and cost (SWaP-C) are increasingly used for intelligence, surveillance, and reconnaissance (ISR) missions due to their affordability, attritability, and suitability for distributed operations~\cite{Replicator_2024,barnett2010optimising}. When deployed in swarms or coordinated teams, they can deliver persistent coverage at lower risk and cost than traditional high-value assets. However, designing effective ISR systems with low-SWaP-C UAVs introduces challenges in endurance, payload capacity, and sensing performance, requiring engineers to evaluate mission-level effectiveness rather than single-platform capability.


Traditional UAV design methods that optimize individual platforms in isolation cannot capture the coupled trade-offs among sensor performance, trajectory geometry, and fleet coordination that drive system-of-systems (SoS) behavior. Addressing these challenges demands a mission engineering framework capable of linking operational design variables such as platform count, sensor configuration, and coordination strategy, to mission outcomes through automated optimization and integrated simulation.


This paper presents such a framework, tailored to the design and evaluation of low-SWaP-C UAV ISR architectures. It unifies design of experiments, multi-objective optimization, and mission-specific simulation into a closed-loop process that embeds estimator-informed performance metrics within the optimization. The approach enables rigorous exploration of trade spaces spanning sensing geometry, team coordination, and cost, while maintaining analytical traceability to mission effectiveness.


The framework is demonstrated through a case study on fixed–field-of-view (FOV) UAVs localizing a man overboard (MOB) in a global navigation satellite system (GNSS)-denied maritime environment.\footnote{The case study in this paper of localizing a MOB in a maritime environment under GNSS-denied conditions is becoming increasingly common due to significant GNSS signal interference in several regions of the world with significant commercial shipping traffic, fishing fleets, and pleasure craft where maritime rescue services routinely operate \cite{GNSS2024}.} The case illustrates how system-level trade-offs between accuracy, affordability, and redundancy can be revealed and optimized quantitatively. Although the scenario focuses on maritime search and rescue, the framework generalizes to other ISR applications, providing a scalable and repeatable tool for early-phase SoS design and mission analysis.


\section{Background}

\subsection{Mission Engineering}
In defense and aerospace, mission engineering provides an operational lens for analyzing how multiple systems deliver mission outcomes. The DoD defines it as an ``interdisciplinary process encompassing the entire technical effort to analyze, design, and integrate current and emerging operational needs and capabilities to achieve desired mission outcomes''~\cite{ME_Guide_2023}. Wertz emphasizes meeting mission objectives at minimum cost and risk~\cite{Wertz_2013}. Practically, mission engineering employs digital models, mission threads, and model-based systems engineering (MBSE) to link measures of performance (MOPs) and measures of effectiveness (MOEs) to mission‐level measures of success (MOSs), enabling quantitative evaluation of alternative concepts of operation (CONOPS)~\cite{ME_Guide_2023}. This study adopts that view, using an operational scenario and simulation to trace design choices directly to mission outcomes.

\subsection{Comparison of Mission Engineering to Systems Engineering}
Traditional systems engineering (SE) emphasizes requirement decomposition, traceability, and subsequent verification \& validation (V\&V) across the system lifecycle. The classic linear or ``Vee'' model of systems engineering ensures traceability from stakeholder needs to design specifications~\cite{Buede_2009,Blanchard_Fabrycky_2011}. Mission engineering complements SE by shifting early-phase attention from delivering a system to achieving mission effects, especially in the SoS contexts where coordination, uncertainty, and evolving constraints dominate~\cite{Hernandez_Pollman_2022}. Whereas SE relies on stabilized baselines and formal change control, mission engineering prioritizes agility via digital mission threads and executable scenarios~\cite{Kennedy_2024}. For low-SWaP-C UAVs, this shift reduces overhead and accelerates convergence on architectures that balance performance, cost, and risk. In practice, mission engineering leads early trades and concept exploration, while SE provides the technical backbone for realization and certification.

\subsection{Mission Engineering Validation Methods}
Classical V\&V generally includes requirements traceability, prototyping, and staged reviews to build stakeholder confidence, however this assumes stable requirements and fixed configurations~\cite{SE_guide_2022,Eng_of_DoD_Sys_2022}. For rapidly evolving and complex SoS, such assumptions break down. Extensive physical testing becomes costly, and suitability metrics may not scale with system size~\cite{Honour_2013,V_and_V_DTIC_2023}. Contemporary practice therefore augments V\&V with model-based and context-driven methods (digital engineering, MBSE, and digital twins) to execute full mission-level simulations, quantify risk, and iterate quickly under uncertainty~\cite{Chaudemar_2024,Dahmann_2021,Mendi_Erol_Doğan_2022,VanBossuyt_2025}. 

These modern mission engineering approaches explicitly tackle the gaps left by the standard systems engineering methods. Where requirements were uncertain or evolving, model-based approaches offer a way to iterate and verify designs amid changing parameters. While physical testing may be too slow or sparse, simulation-based testing provides speed and scale. By validating not only that a system meets its specifications, but that it can accomplish its mission effectively under various conditions, modern mission engineering techniques dramatically enhance V\&V for autonomous and complex aerospace and defense systems.

\section{Literature Review}
Recent research in mission engineering emphasizes a shift from platform-centric to mission-centric analysis, particularly for small, low-cost UAVs operating in collaborative teams where success depends on SoS-level orchestration rather than individual capability. Mission-based frameworks using MBSE have been proposed to model swarm behavior and evaluate mission outcomes through MOPs and MOEs~\cite{Giles_Giammarco_2019}. Galvez-Serna \emph{et al.} extend this by demonstrating a modular, software-in-the-loop architecture for rapid transition from simulation to field deployment~\cite{Galvez_2022}. Such approaches are directly relevant to low-SWaP-C UAV ISR systems, where trajectories and sensor performance must be validated in simulation prior to deployment.


The integration of optimization within simulation-based mission design has become a key research direction. Programs such as DARPA CODE, AFRL Golden Horde, and ONR LOCUST emphasize cooperative autonomy for contested environments~\cite{DARPA_CODE_2019,AFRL_2021,LOCUST_2015}, while academic studies address fleet composition and trajectory optimization using evolutionary and hybrid algorithms~\cite{Bociaga_Crossley_2013,Panowicz_Stecz_2024,Liu_2022}. These efforts underscore the importance of embedding search algorithms directly within mission-level simulation to manage complex multi-dimensional trade spaces.


Another major theme is sensor architecture trade-offs for expendable ISR platforms. Gimballed cameras provide sensor agility but impose penalties in complexity, power, and structural demands, raising acquisition and sustainment costs~\cite{Keane_Sobester_Scanlan_2017, Korobov_Shipitko_Konovalenko_Grigoryev_Chukalina_2020}. By contrast, fixed-FOV architectures enable lighter UAVs, reduced cost, and scalability for swarms~\cite{Telli_2023, Li_2025}. These benefits are offset by planning complexity, since trajectory design must explicitly account for FOV geometry and line-of-sight constraints~\cite{Zhou_2024}. Nevertheless, fixed-FOV systems improve endurance, reduce lifecycle cost~\cite{Miller_Mooty_Hilkert_2013, Hovenburg_Johansen_Storvold_2017}, leverage commercial off the shelf (COTS) components~\cite{rodin_survey_2019, li_optimal_2023}, and support swarm-level redundancy and graceful degradation~\cite{Petritoli_Leccese_Ciani_2018,Liu_2022}.

Despite this progress, current research remains fragmented. UAV design, trajectory optimization, system architecture, and mission simulation are often pursued independently, limiting cross-domain insight~\cite{VanBossuyt_2019}. Persistent gaps include: (i) lack of joint optimization of platform and trajectory under realistic constraints (e.g., communication limits, no-fly zones, GNSS denial); (ii) weak integration of cost and reliability with estimator-driven design; and (iii) limited unification of digital simulation, system synthesis, and multi-objective optimization under uncertainty. This paper addresses these gaps through a unified, modular mission-engineering framework that automates design exploration and evaluation of system architectures. While demonstrated for low-SWaP-C ISR UAVs, the framework generalizes to other SoS missions where platform configuration, sensing strategies, and trajectory planning must be tightly integrated.

\section{Mission Engineering Framework}

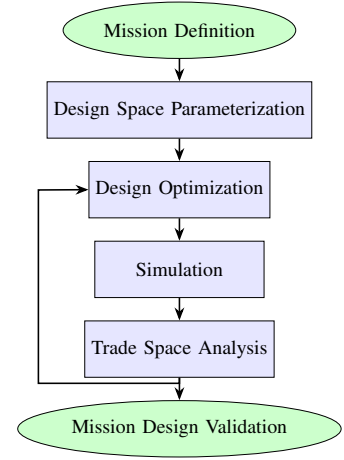
\begin{figure}
\begin{center}
\scalebox{0.75}{
\begin{tikzpicture}[node distance=0.4cm and 0.4cm]

\node (start) [startstop] {Mission Definition};
\node (trade) [process, below=of start] {Design Space Parameterization};
\node (opt) [process, below=of trade] {Design Optimization};
\node (sim) [process, below=of opt] {Simulation};
\node (data) [process, below=of sim] {Trade Space Analysis};
\node (end) [startstop, below=of data] {Mission Design Validation};

\draw [arrow] (start) -- (trade);
\draw [arrow] (trade) -- (opt);
\draw [arrow] (opt) -- (sim);
\draw [arrow] (sim) -- (data);
\draw [arrow] (data) -- (end);

\draw [arrow] (data.south) --++ (0,-0.1) --++ (-2.5,0) --++ (0,3.43) -- (opt.west);



\end{tikzpicture}}
\end{center}
\caption{Mission engineering framework methodology. This high-level diagram depicts the automated process used in this study to determine the optimized system architecture based on mission requirements, constraints, and outcomes.}
\label{fig:Mission Engineering Framework}
\end{figure}

The foundation of this work is the development of a repeatable, metric-driven mission engineering framework for agile system design, by embedding a formal design of experiments (DOE) approach within the mission engineering process~\cite{Maier_Rechtin_2009}. The resulting framework depicted in Fig.~\ref{fig:Mission Engineering Framework}, is designed to efficiently explore the architecture trade space while accounting for operational constraints such as sensor limits, communication bounds, and mission boundaries. Rather than optimizing individual components in isolation, the framework evaluates configurations holistically at the mission level, recognizing that the optimal SoS solution may entail suboptimal individual designs in favor of overall SoS effectiveness.

This integration of DOE enables systematic exploration of high-dimensional, nonlinear design spaces by combining structured sampling strategies with multi-objective optimization. Latin hypercube sampling (LHS) is employed to initialize the design space exploration with representative coverage, ensuring that early-stage evaluations capture the breadth of system sensitivities. This is followed by a multi-objective optimization using a genetic algorithm (GA), which iteratively evolves candidate solutions based on a defined fitness function. The approach is particularly suited to problems characterized by discrete architectural choices, complex mission constraints, and non-convex objective landscapes.

Each candidate design is evaluated using mission-specific simulations that capture vehicle kinematics, sensor performance, mission boundaries, and communication constraints. The simulation outcomes are used to compute mission-level metrics, such as the defined MOEs and MOPs. These results inform both the selection of optimal architectures and the underlying sensitivity analysis, enabling decision-makers to understand the influence of design variables on system outcomes. By extracting the Pareto front from the final design population, the framework reveals the trade space between competing objectives such as mission effectiveness and sensor geometry. This trade space allows mission designers to identify configurations that best balance performance within operational constraints.

Through this process, the methodology provides a unified structure for system exploration, optimization, and validation. Although demonstrated in the context of UAV-based ISR in GNSS-denied environments, the framework generalizes to any SoS problem where coordinated platform behavior, sensing constraints, and real-world limitations must be jointly considered. 

\subsection{Mission Definition}
The primary motivation is to enable rapid evaluation and design of performance-driven systems suitable for dynamic operational environments. This capability necessitates a clear definition of mission objectives, operational context, and environmental constraints.  These include parameters such as mission outcomes, target behavior models, geographic and operational boundaries, and sensing modalities. Collectively, these inputs inform the broader set of capability needs that the system must address. 

The mission definition serves as the foundation for the entire engineering process by identifying the MOEs and MOPs that will be used to evaluate candidate systems. For the design to be verifiable and traceable, these MOEs and MOPs must directly link to the specific SoS or system functions they are measuring,  which themselves are derived from the requirements established during mission definition. A critical aspect of this traceability is the emphasis on SWaP-C prioritization, which ensures that the resulting architectures are not only operationally effective but also feasible for rapid fielding in support of mission objectives. 

Scenario modeling formalizes these elements into a simulation-ready format, translating conceptual mission constructs into executable models. This includes the placement and behavior of adversarial elements, delineation of contested or denied areas, and specifications for initial vehicle deployment. The output of this step defines the operational constraints and performance benchmarks that all candidate solutions must satisfy.

\subsection{Design Space Parameterization}
Once the mission environment is established, the next phase involves defining the design variable space. This includes architectural decisions such as the number and type of vehicles, sensor configurations, communication requirements, and coordination strategies. Each variable is bounded based on feasibility and relevance to mission outcomes. Design variables may be discrete (e.g., number of vehicles, vehicle type) or continuous (e.g., trajectory parameters, mission duration). These variables form the input structure for the DOE process and represent the design variables through which system-level trade-offs are explored.

To efficiently populate the high-dimensional design space with diverse and representative samples, LHS is applied~\cite{McKay_Beckman_Conover_1979}. This ensures broad coverage without redundancy, enabling early identification of system sensitivities and avoiding premature convergence to local optima. LHS selects near-random design points across each dimension, making it well suited to capturing nonlinear interactions in system performance. These initial samples serve as the input to early-stage simulation, allowing the framework to map the relationship between design inputs and mission outcomes.

\subsection{Design Optimization}
A GA is employed to refine the solution space based on the designs developed from the design space parameterization~\cite{Petrowski_Ben-Hamida_2017}. Candidate solutions are evaluated and ranked based on a fitness function that reflects a weighted combination of mission objectives. GAs use biologically inspired operations—selection, crossover, mutation—to evolve new designs over successive generations. This enables the discovery of high-performing configurations in complex, non-convex search spaces with discrete architectural choices and nonlinear mission constraints. The optimization identifies Pareto-optimal solutions that simultaneously improve multiple conflicting objectives.

\subsection{Simulation}
Each sampled design is passed to a mission-specific simulation, such as a digital twin, that models individual system and SoS dynamics, sensor excitation, communication behavior, and environmental influences as applicable. This simulation evaluates the effectiveness of each candidate in terms of defined MOEs and MOPs such as localization accuracy, target coverage duration, communication link availability, and endurance. This evaluation step is tightly integrated with the mission model and enforces operational constraints such as exclusion zones, sensor constraints, and system dynamics limits. The simulation results quantify system-level effectiveness and form the basis for the design optimization and trade space analysis.

\subsection{Trade Space Analysis}
The candidate designs generated by the GA are analyzed to extract the Pareto front, revealing the trade space between mission performance and design cost or complexity. Sensitivity analysis is conducted to determine which variables most influence mission outcomes, providing insight into the robustness and adaptability of the architecture. This information supports both down-selection of viable system configurations and further refinement of constraints for future design iterations. 





\subsection{Mission Design Validation}
The Validation Square, as proposed by Pedersen et al.~\cite{Pedersen__2000}, provides a structured framework for evaluating the usefulness of design methods through both qualitative and quantitative measures. It assesses design validity across theoretical, empirical, structural, and practical dimensions, offering a balanced approach to evaluating methodological approaches. This validation model is particularly well-suited for mission engineering, where complex systems are developed in uncertain and dynamic operational contexts. Unlike traditional validation frameworks grounded solely in logical deduction or empirical testing, the Validation Square recognizes that engineering design often involves open-ended problems, subjective judgments, and heuristic-based methods. All of these are common conditions in mission-level planning and architecture development.

Candidate architectures on or near the Pareto front are selected based on mission priorities, available resources, and operational constraints. These configurations may undergo further validation through repeated simulations with scenario variations (e.g., target maneuverability, mission boundaries, communication degradation) to ensure robustness. In practice, selected designs may be exported for hardware-in-the-loop or software-in-the-loop testing, enabling integration with downstream acquisition workflows. This final step confirms that the selected architecture meets mission requirements under realistic operational conditions.

\section{Methods and Case Study: Application of the Mission Engineering Framework}

This section demonstrates the application of the proposed mission engineering framework to a representative bearing-only ISR problem involving low-SWaP-C UAVs equipped with fixed-FOV electro-optical sensors operating in a GNSS-denied maritime environment. The case study serves to illustrate how the framework integrates estimator-informed performance evaluation, trajectory optimization, and cost modeling into a unified, simulation-driven design process. The implementation enables automated exploration of candidate architectures by embedding an estimator and vehicle simulation within a multi-objective optimization loop, thereby linking system performance and affordability in a single decision space. 

The framework is implemented as an automated MATLAB-based simulation and optimization suite capable of batch-processing large design sets generated through LHS. The workflow proceeds through six stages: mission definition, design space parameterization, design optimization, simulation, trade space analysis, and validation. Together, these stages form a closed-loop process that systematically maps design decisions to mission outcomes. A case study focused on \emph{camera angle, camera resolution}, and \emph{team size}, is presented to illustrate the methodology and demonstrate how cost-aware optimization reveals the balance between estimator accuracy, mission affordability, and architectural resilience.

\subsection{Mission Definition}
\begin{figure}
    \centering
    \includegraphics[width=0.8\linewidth]{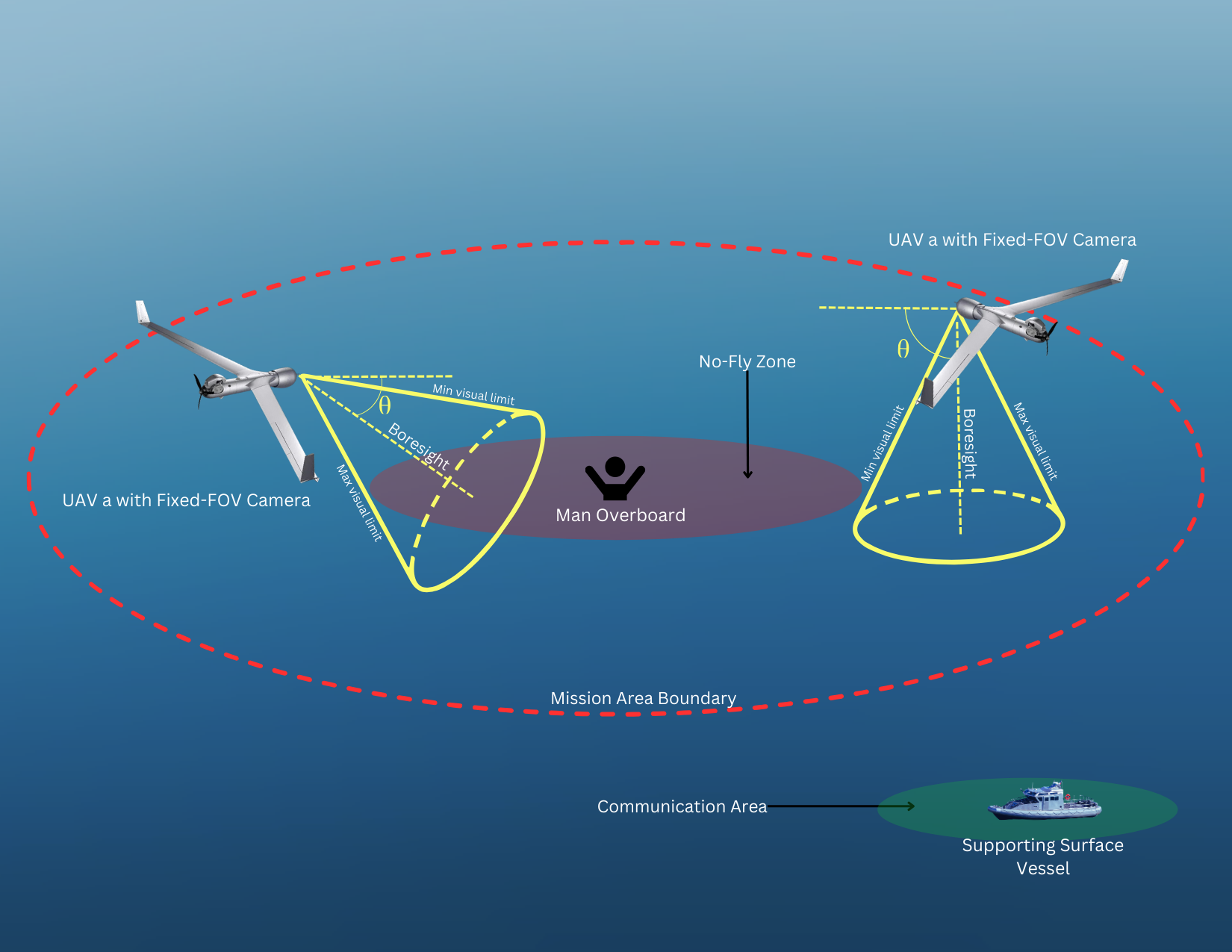}
    \caption{Concept of Operations for UAV-based MOB localization. The left UAV achieves successful localization because the MOB remains within its camera boresight, while the right UAV fails due to its boresight misalignment. This geometry highlights the critical role of sensor angle in determining both observability and the feasible trajectories UAVs can adopt to acquire valid bearing measurements and support mission success.}
    \label{fig:CONOP}
\end{figure}

The representative mission illustrated in Fig.~\ref{fig:CONOP} involves small, low-SWaP-C UAVs deployed from a supporting vessel to localize a MOB using passive electro-optical sensors. The operational context assumes a GNSS-denied and electromagnetically contested environment, requiring passive sensing and return-to-communicate strategies. UAVs operate at fixed altitude with three principal constraints: (i) a no-fly safety buffer around the estimated MOB location to reduce collision risk, (ii) a bounded communication radius with the supporting surface vessel, and (iii) a predefined mission area based on a MOB area of uncertainty region based on set-and-drift.~\footnote{In maritime navigation, ``set” refers to the direction toward which a current or tide moves a vessel, while ``drift” denotes the rate or speed of that movement, typically expressed in knots. Together, set-and-drift describe the vector displacement caused by water motion relative to the Earth, which must be accounted for when predicting the position of objects or vehicles affected by currents\cite{Bowditch}.}  

The primary MOEs for the scenarios are the root-mean-square errors (RMSE) for each target generated by the embedded simulation. MOPs include adherence to communication, no-fly zone, and mission area constraints.  


\subsection{Design Space Parameterization}
The design space integrates both discrete and continuous variables that govern the architectural and trajectory-level design of the UAV system. Discrete parameters include the number of UAVs and the sensing configuration, specifically, the camera resolution. Continuous parameters include the camera boresight angle and trajectory control points, defined in a Bernstein polynomial basis to ensure smooth, dynamically feasible flight paths under bank-to-turn kinematics. The Bernstein basis provides a numerically stable and convex-hull–bounded representation that enforces trajectory smoothness and boundary conditions~\cite{Mudrik_Sciacchitano_Kragelund_Kang_Kaminer_2025}.

For this case study, the design space jointly sweeps three primary factors: team size ($N_{\text{UAV}}$), camera resolution, and camera boresight angle. Specifically, team size is restricted to $N_{\text{UAV}}\in\{1,3\}$, to bound runtime while preserving decision relevance. Camera resolution is selected from a cataloged set of commercially available sensors, each defined by horizontal and vertical resolution, as well as total megapixels (MP). Boresight angle $\theta_b$ varies from $0^{\circ}$  to $90^{\circ}$ in north-east-down coordinates. LHS is used to generate an initial population that uniformly spans these dimensions, ensuring independent variation of each variable and capturing nonlinear interactions between geometry, sensing performance, and cost.  

\subsection{Design Optimization}
A variant of the Non-dominated Sorting Genetic Algorithm II (NSGA-II), a widely used multi-objective evolutionary optimization method, is employed to evolve and refine the initial population of designs based on the mission-specific objectives~\cite{Deb_2002}. The fitness of each design is evaluated through simulation outputs that quantify system-level trade-offs between estimator performance, mission cost, and architectural redundancy. As the GA iterates over generations, new candidate architectures are generated and evaluated, allowing the algorithm to converge toward Pareto-optimal solutions in the high-dimensional, non-convex trade space. The resulting Pareto set yields \emph{feasible trajectories} used to localize both targets. These trajectories are derived from combinations of camera resolution, boresight angle, and team size that balance localization accuracy, affordability, and robustness.

To explicitly incorporate affordability into the optimization, the total mission cost is defined as
\begin{equation}
    C_{\text{mission}} \;=\; N_{\text{UAV}} \big( C_{\text{platform}} + C_{\text{sensor}} \big) \;+\; C_{\text{support}},
\end{equation}
\noindent where $N_{\text{UAV}}$ is the number of deployed vehicles, $C_{\text{platform}}$ is the per-airframe unit cost, $C_{\text{sensor}}$ represents the payload (camera) cost, and $C_{\text{support}}$ accounts for auxiliary mission activities such as logistics and launch support. This formulation embeds cost as a continuous design objective that scales with both team size and sensor capability. The assumed costs used in this study are $\$500$ (USD) per UAV platform and $\$0$ (USD) for support.

The sensor cost term $C_{\text{sensor}}$ is derived from a data-driven model linking payload price directly to imaging performance. Each camera configuration is parameterized by its horizontal and vertical resolution $(r_x, r_y)$ and corresponding total MP count,
\begin{equation}
    \text{MP} = \frac{r_x r_y}{10^6}.
\end{equation}
An empirical relationship, obtained from market data across a large set of COTS cameras, defines the payload cost as
\begin{equation}
    C_{\text{sensor}}(\text{MP}) = k_{\text{res}} \, \text{MP},
\end{equation}
\noindent where $k_{\text{res}} = 12.698~\$/\text{MP}$ represents the fitted slope of the resolution–price curve.~\footnote{The price–resolution relationship was obtained from a linear regression of camera module data downloaded directly from the Digi-Key catalog in October 2025. Details of the dataset and analysis scripts are provided in Appendix~\ref{app:Camera_specs}.} This linear model captures the proportional increase in payload cost with sensor resolution, allowing the optimizer to trade optical capability directly against mission expenditure. Higher-resolution sensors enhance imaging fidelity and improve FOV coverage, but they also increase mission cost. By linking $C_{\text{sensor}}$ to $\text{MP}$, the optimization embeds a continuous affordability gradient that drives the search toward cost-effective combinations of camera resolution and platform count.  

During simulation, each candidate design passes a structured camera variable containing $(r_x, r_y, \text{MP})$ and the corresponding team size $N_{\text{UAV}}$. The resolution parameters $(r_x, r_y)$ are used within the sensor-geometry model to compute instantaneous FOV, while $\text{MP}$ and $N_{\text{UAV}}$ enter the cost model to evaluate $C_{\text{mission}}$. This coupling allows the GA to explore the joint design space of geometric observability, sensing fidelity, and cost efficiency, ensuring that affordability and redundancy are treated as intrinsic design drivers rather than post-hoc evaluation metrics.


Because NSGA-II natively handles multiple objectives, the GA minimizes a vector objective composed of the Monte Carlo–averaged RMSE for each target and the mission cost,

\begin{equation}
    \min_{\boldsymbol{\theta}} \;\; 
    \mathbf{f}(\boldsymbol{\theta}) 
    \;=\; 
    \begin{bmatrix}
        \overline{\mathrm{RMSE}}_{\text{MOB}}(\boldsymbol{\theta})\\[2pt]
        \overline{\mathrm{RMSE}}_{\text{vessel}}(\boldsymbol{\theta})\\[2pt]
        C_{\text{mission}}(\boldsymbol{\theta})
    \end{bmatrix},
\end{equation}
\noindent where $\boldsymbol{\theta}$ collects the decision variables (camera boresight angle, sensor resolution, and UAV team size) and the overline denotes an average over Monte Carlo realizations. This formulation allows the optimizer to discover Pareto-optimal architectures that balance estimation accuracy, affordability, and robustness across the full design space.

\subsection{Simulation}
Each candidate design is evaluated within a high-fidelity simulation environment that models UAV kinematics, target dynamics, sensor geometry, and estimator performance. This environment, embedded within the DOE and optimization framework, captures the nonlinear coupling between platform motion, FOV constraints, and estimator accuracy under stochastic noise. The simulation thus forms the analytical backbone of the mission engineering framework, mapping each candidate design to its resulting MOE metrics of localization RMSE for both the MOB and the supporting vessel.

The simulation scales with the number of UAVs specified by the design vector, with each platform governed by bank-to-turn kinematics that reflect the nonholonomic constraints of fixed-wing flight. UAVs maintain constant altitude and are controlled via roll-angle commands that induce lateral acceleration and heading change. Sensors are modeled as passive, forward-facing electro-optical cameras producing noisy azimuth and elevation measurements relative to the target. Because the cameras are fixed in the UAV body frame, valid bearing measurements are only obtained when the target lies within the FOV, enforcing directional observability consistent with real-world operations.

For each candidate configuration, feasible UAV trajectories are generated by solving a trajectory optimization problem that minimizes an information-theoretic cost function
\begin{dmath}
         \min_{\mathbf{x}(t),\mathbf{u}(t)} \quad J = w_1 \cdot \mathrm{Tr}(\mathrm{PCRLB}_{\text{MOB}})\\
      + (1-w_1)\cdot \mathrm{Tr}(\mathrm{PCRLB}_{\text{vessel}})
      + w_2 \cdot t_f
\end{dmath}
\noindent where $\mathbf{x}(t)$ and $\mathbf{u}(t)$ are the state and control trajectories for all UAVs over time. The Posterior Cram\'er-Rao Lower Bound (PCRLB) is an information-based proxy for estimator performance that is widely used in trajectory optimization and is the theoretical lower bound on the covariance of any unbiased estimator~~\cite{nardone_fundamental_1984,hammel_optimal_1989,passerieux_optimal_1998, Tichavsky_1998, oshman_optimization_1999,grocholsky_information-theoretic_2003,ousingsawat_optimal_2007,levine_information-theoretic_2013,lee_cooperative_2013, VanTrees_2013,pedro_underwater_2015,vander_hook_algorithms_2015,ponda_trajectory_2009}. $\mathrm{Tr}(\cdot)$ denotes the matrix trace operator, and $t_f$ is the total mission duration. The weights $w_1$ and $w_2$ tune the tradeoff between estimation uncertainty and operational time.

The PCRLB is propagated using the Fisher Information Matrix (FIM), which is updated only when the target is within the camera’s FOV, consistent with established information-based trajectory planning methods, where the $\mathrm{Tr}(\text{PCRLB})$ serves as a proxy for observability~\cite{ponda_trajectory_2009}. 
This formulation links trajectory generation directly to estimator observability, ensuring geometric feasibility and information efficiency.

The optimization is subject to several dynamic and mission-related constraints. These include UAV kinematics, UAV bank-angle bounds, spatial boundary limits imposed by set-and-drift, MOB avoidance enforced via a no-fly-zone radius ($\text{r}_{\text{NFZ}}$), and return-to-communicate requirements ($\text{r}_{\text{com}}$):
\begin{subequations}
\begin{align}
    &\dot{\mathbf{x}}_{i}(t) = f(\mathbf{x}_{i}(t), \mathbf{u}_{i}(t)), \\
    &\phi^{\min} \leq \phi_{A,i}(t) \leq \phi^{\max}, \quad &\forall\, t \in [0, t_f], \\
    &x_{\min} \leq x_{A,i}(t) \leq x_{\max}, \quad &\forall t \in [0, t_f], \\
    &y_{\min} \leq y_{A,i}(t) \leq y_{\max}, \quad &\forall t \in [0, t_f], \\
    &\|\mathbf{p}_{MOB}(t) - \mathbf{p}_{A,i}(t)\| \geq r_{\text{NFZ}}, \quad &\forall\, t \in [0, t_f], \\
    &\|\mathbf{p}_{\text{vessel}}(t_f) - \mathbf{p}_{A,i}(t_f)\| \leq r_{\text{com}}
\end{align}
\end{subequations}

\noindent where $\mathbf{p}_{A,i}(t)$ is the position of UAV $i$ at time $t$, $\phi_A(t)$ is the roll angle control input, and the subscripted limits enforce both trajectory feasibility and operational compliance.

Once feasible trajectories are generated, Monte Carlo simulations evaluate estimator performance across noise realizations. Each UAV runs an onboard extended Kalman filter (EKF) to track both the MOB and the supporting vessel using bearing-only measurements. At mission completion, when UAVs reestablish communication with the supporting vessel, a federated Kalman filter (FKF) fuses their local estimates using an information-weighted consensus rule. This fusion produces a global posterior estimate of all target states, enabling assessment of overall mission accuracy. For single-UAV configurations, the FKF solution reduces to the EKF result.

The estimator is executed across multiple stochastic trials to compute Monte Carlo–averaged RMSE values for both targets. These results are returned to the GA's optimization loop as quantitative measures of each design’s performance. In this way, the simulation environment directly couples sensing, control, and estimator dynamics with system cost and architecture, providing a rigorous foundation for the trade space analysis that follows.

 \subsection{Trade Space Analysis}
The simulation results from all evaluated candidate designs are aggregated to construct the trade space. Each design is evaluated against the mission MOPs, and any trajectory that violates no-fly zones, communication limits, or mission boundaries is automatically discarded. The remaining feasible designs are ranked according to their mission-level MOEs, specifically the Monte Carlo–averaged RMSE for the MOB and the supporting vessel. These metrics form the basis for extracting Pareto fronts that describe the competing relationships between estimator performance, sensing geometry, and cost.  

\begin{figure}[!htb]
    \centering
     \includegraphics[width=\linewidth,height=0.4\textheight,keepaspectratio]{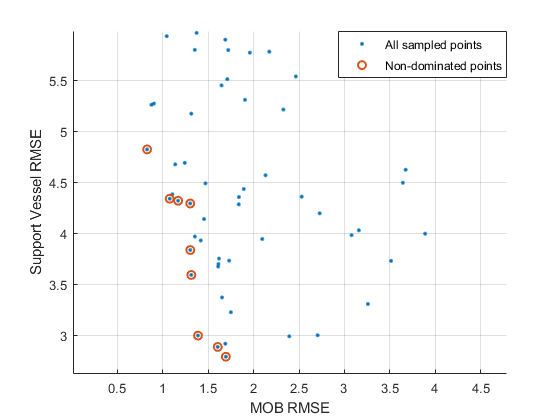}
    \caption{Pareto front obtained from optimizing UAV camera elevation angles between 0 and 90 degrees (NED coordinates) at a fixed 50 m altitude for the single-UAV configuration. The x-axis represents localization accuracy for the MOB while the y-axis shows the competing localization accuracy of the supporting vessel.} 
    \label{fig:Pareto_Camera_Angle}
\end{figure}

\begin{figure}[!htb]
      \centering
       \includegraphics[width=\linewidth,height=0.4\textheight,keepaspectratio]{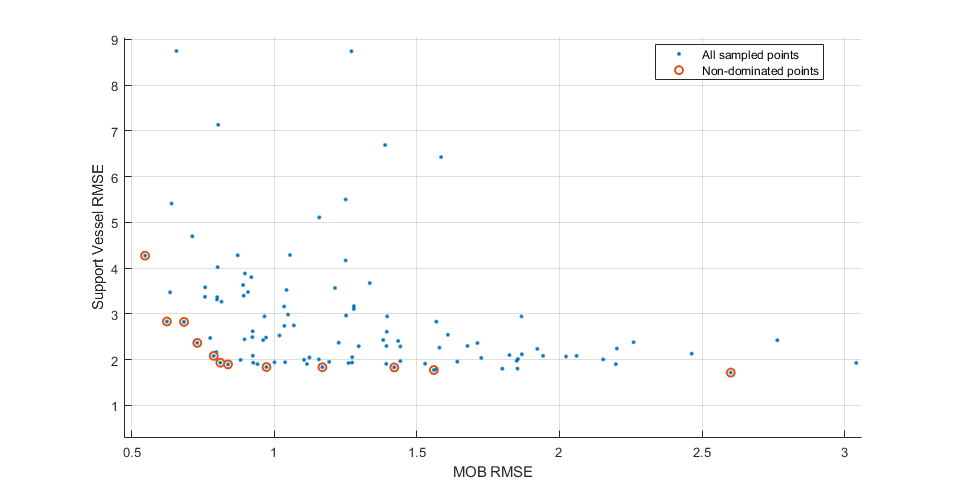}
      \caption{Pareto front obtained from optimizing UAV camera elevation angles between 0 and 90 degrees (NED coordinates) at fixed 50 m, 60 m and 70 m altitudes for a team of three UAVs, respectively. The x-axis represents localization accuracy for the MOB while the y-axis shows the competing localization accuracy of the supporting vessel.} 
      \label{fig:Pareto_Camera_Angle_multi}
  \end{figure}

Figures~\ref{fig:Pareto_Camera_Angle} and~\ref{fig:Pareto_Camera_Angle_multi} illustrate how the GA evolves successive design generations to identify non-dominated solutions for different UAV configurations. In this example, both the single- and three-UAV systems use identical camera resolutions but vary in boresight angle. The GA’s operation is visualized through the formation of Pareto fronts, which represent the locus of optimal trade-offs between MOB and vessel localization accuracy.  

For the single-UAV case illustrated in Fig.~\ref{fig:Pareto_Camera_Angle}, a sweep over camera angles produces a Pareto frontier in RMSE-space. As expected, improving localization accuracy for the MOB comes at the expense of accuracy for the supporting vessel, since the UAV must allocate its limited sensing footprint between the two targets.  The absence of a single angle that minimizes both errors highlights the geometric trade-offs inherent in bearing-only sensing with fixed-FOV sensors. Table~\ref{tab:pareto_solutions} summarizes the Pareto-optimal camera angles for this case.
 
\begin{table}[!htb]
\caption{Pareto-Optimal Camera Angles and RMSE (Single-UAV)}
\label{tab:pareto_solutions}
\centering
\begin{tabular}{ccc}
\toprule
\textbf{Camera Angle ($^\circ$)} & $\mathbf{RMSE_{MOB}}$ (m) & $\mathbf{RMSE_{vessel}}$ (m) \\
\midrule
33.0 & 1.08 & 4.34 \\
34.5 & 1.39 & 2.99 \\
37.4 & 1.70 & 2.78 \\
41.1 & 1.31 & 3.60 \\
44.3 & 1.30 & 3.83 \\
46.6 & 1.60 & 2.88 \\
46.6 & 1.20 & 4.32 \\
46.7 & 1.30 & 4.29 \\
48.2 & 0.83 & 4.82 \\
\bottomrule
\end{tabular}
\end{table}

In contrast, the three-UAV configuration in Fig.~\ref{fig:Pareto_Camera_Angle_multi} yields a more compact Pareto front, particularly along the MOB axis. This results from the increased geometric diversity provided by multiple observers, which enhances overall observability and reduces sensitivity to boresight selection. Of note, the optimal boresight angles for a team of UAVs coordinating to localize the MOB and supporting vessel do not match the angles used by a single UAV, highlighting the need for joint design optimization strategies as is employed here. The clustering of optimal angles near $34^{\circ}$ indicates that exact precision in angle placement is not necessary; instead, engineering tolerances can be relaxed to allow for small adjustments in angle as a function of flight altitude or mission phase. This flexibility reduces design complexity while maintaining estimator performance. Table~\ref{tab:pareto_solutions_multi} lists the corresponding Pareto-optimal values for this multi-UAV scenario.

\begin{table}[!htb]
\caption{Pareto-Optimal Camera Angles and RMSE (Multi-UAV)}
\label{tab:pareto_solutions_multi}
\centering
\begin{tabular}{ccc}
\toprule
\textbf{Camera Angle ($^\circ$)} & $\mathbf{RMSE_{MOB}}$ (m) & $\mathbf{RMSE_{vessel}}$ (m) \\
\midrule
6.5  & 2.60 & 1.71 \\
29.6 & 1.56 & 1.77 \\
29.6 & 0.62 & 2.83 \\
33.2 & 0.55 & 4.27 \\
33.8 & 0.81 & 1.93 \\
34.1 & 1.17 & 1.83 \\
34.2 & 0.84 & 1.89 \\
34.2 & 1.42 & 1.83 \\
34.4 & 0.97 & 1.84 \\
34.4 & 0.68 & 2.82 \\
34.5 & 0.79 & 2.08 \\
43.7 & 0.73 & 2.36 \\
\bottomrule
\end{tabular}
\end{table}

\begin{figure}
    \centering
    \includegraphics[width=0.9\linewidth]{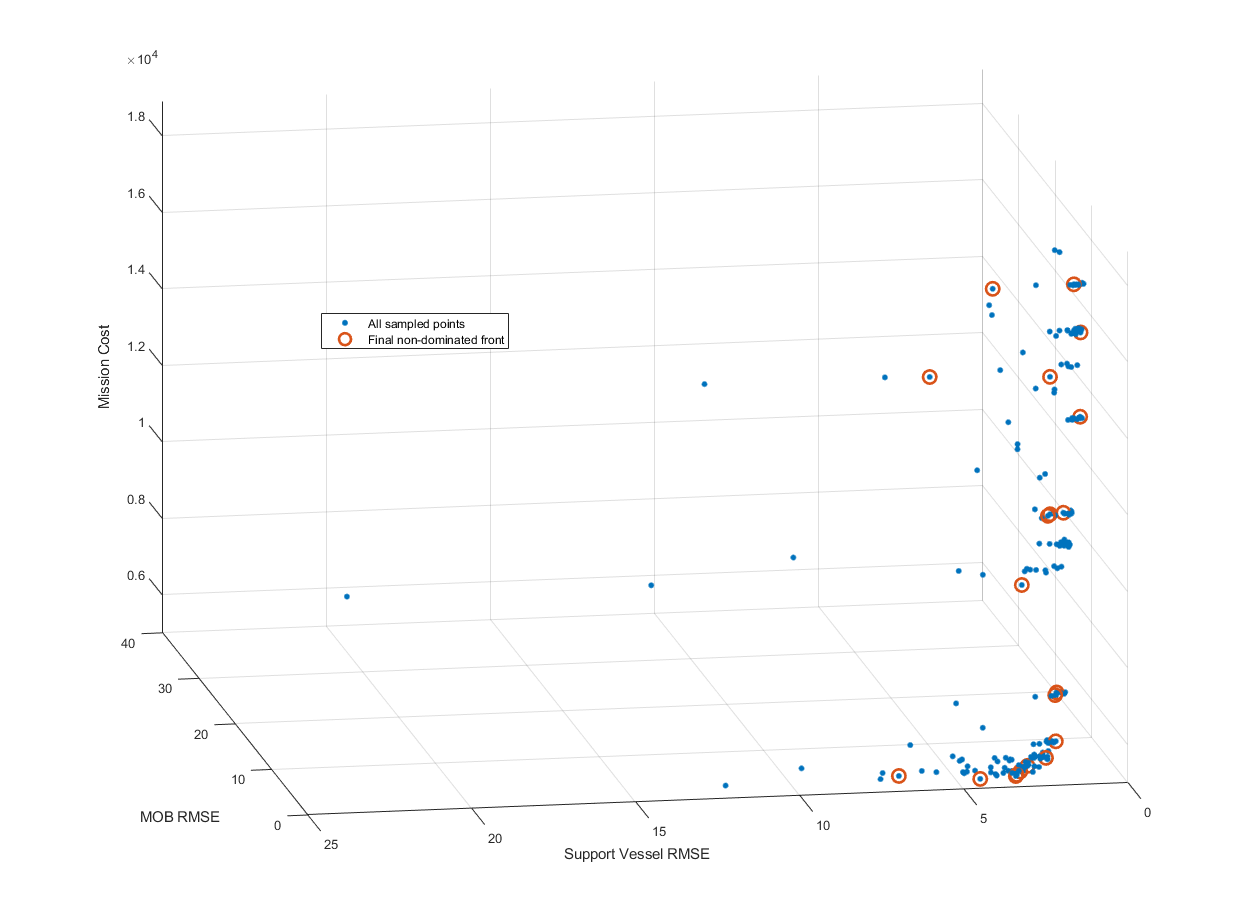}
    \caption{Three-dimensional Pareto front showing the relationship between camera boresight angle, number of UAVs, and total mission cost. Each point represents a candidate design evaluated through Monte Carlo simulation, with red markers denoting the Pareto-optimal solutions.}
    \label{fig:Pareto_3D}
\end{figure}

The final output for use by the mission engineer incorporates mission cost as a co-optimized objective to reveal system-level affordability trends. Figure~\ref{fig:Pareto_3D} presents the three-dimensional Pareto front relating MOB and vessel localization accuracy to total mission cost. To aid visualization of the three-dimensional Pareto front, two-dimensional projections of the trade space are provided in Appendix~\ref{app:ParetoProjections}. Each point corresponds to a unique combination of UAV count, boresight angle, and camera resolution evaluated through Monte Carlo simulation, with red markers denoting non-dominated solutions. Across all architectures, MOB localization accuracy remains below 1~meter, indicating that even the least costly configurations achieve near-optimal performance for this target. Additional investments in team size or sensor resolution produce only marginal improvements for localization of the supporting vessel.

The flattening of the Pareto surface along the MOB axis confirms that beyond a certain threshold, increases in cost primarily yield  redundancy and mission resilience rather than improved accuracy, highlighting where further investment offers diminishing returns. The resulting trade space provides a quantitative basis for selecting architectures that balance estimation precision, affordability, and operational resilience, supporting evidence-based design decisions for low SWaP-C ISR missions.

\subsection{Mission Design Validation}

The final phase of the methodology applies the Validation Square framework to assess the credibility, robustness, and mission relevance of the proposed design approach in the context of MOB recovery. Candidate designs located on or near the Pareto front are selected for human-in-the-loop review, ensuring that numerical optimization outputs align with operational expectations and system-level feasibility.  

Validation is grounded in both theoretical and empirical measures. Theoretical structural and performance validity are established through the generation of dynamically feasible trajectories and the use of PCRLB-based optimization as a benchmark for estimator performance. Empirical validity is confirmed through Monte Carlo trials of the estimator, demonstrating consistent RMSE convergence under measurement noise and scenario uncertainty. Structural realism is achieved through simulation of bank-to-turn vehicle dynamics and fixed-FOV electro-optical sensing models representative of reconnaissance-class UAVs operating under GNSS-denied conditions.  

The validation process also extends beyond estimator performance to include cost and redundancy considerations. Comparative analysis verifies that cost trends predicted by the embedded mission-cost model align with expected scaling relationships across team sizes and sensor configurations. Designs that fall within a feasible region of the cost–performance surface are reviewed for operational practicality, confirming that the most cost-effective architectures also provide sufficient redundancy for mission resilience. This ensures that the framework’s affordability predictions are not purely algorithmic, but remain consistent with engineering judgment and system deployment realities.

Additional validation incorporates comparison to heuristic search-and-rescue patterns and mission-engineering subject-matter expertise. Theoretical considerations assessed include the relative localization performance of single- versus multi-UAV architectures and the degree to which optimized trajectories achieve RMSE convergence toward the PCRLB. Empirical considerations include the tactical feasibility of resulting flight profiles, adherence to safety buffers around the MOB, communication range with the supporting vessel, and compliance of the resulting sensing geometries with accepted ISR and search-and-rescue practices.  

Together, these results satisfy the objectives of the Validation Square. The framework produces outputs that are theoretically justified, empirically consistent, and operationally credible. Most critically, validation confirms that the inclusion of cost as a design variable does not compromise mission performance but rather improves system-level decision quality, revealing affordable, redundant, and dynamically feasible architectures that enhance the probability of successful and timely MOB localization.

\section{Discussion: System-Level Trade-Offs}
The results highlight both the system-level trade-offs in sensing geometry and the broader value of the proposed mission engineering framework as a decision-support tool. The study shows that UAV-based localization systems operating with fixed-FOV, bearing-only sensors face distinct design challenges stemming from restricted kinematics and limited sensor agility. Traditional platform-centric methods cannot capture the emergent SoS behaviors that arise when multiple UAVs coordinate in GNSS-denied environments. By embedding estimator-informed performance metrics directly within the optimization loop, the framework exposes not only the Pareto-optimal configurations but also the sensitivities of mission outcomes to design parameters such as boresight angle, sensor resolution, and team size.

A key finding is the strong contrast between the single- and multi-UAV configurations. In the single-UAV case, the Pareto-optimal camera angles span roughly $33^{\circ}$–$50^{\circ}$ (Table~\ref{tab:pareto_solutions}), reflecting the geometric tradeoff between MOB and vessel localization. The broad distribution underscores the limitations of bearing-only sensing when one platform must balance multiple objectives. By contrast, the multi-UAV optimization converges tightly near $34^{\circ}$ (Table~\ref{tab:pareto_solutions_multi}), demonstrating that distributed sensing geometry provides both higher accuracy and greater robustness to small perturbations in boresight alignment. This convergence highlights the advantage of SoS coordination, where geometric diversity across observers enables simultaneous improvements in both objectives. This confirms that SoS coordination achieves performance levels unattainable through isolated platform optimization, reinforcing the importance of joint design across vehicles.

An additional implication of this convergence is that exact boresight precision is not critical. Small deviations on the order of one degree around the $34^{\circ}$ cluster do not materially degrade performance. This suggests that engineering tolerances for sensor mounting can be relaxed, reducing mechanical complexity and cost while increasing robustness. Identifying these ``robust zones'' of the design space is a key design outcome obtained by the use of this framework, as it points to opportunities for reducing complexity, and subsequently cost, while preserving mission effectiveness.  Moreover, it enables potential in-flight adaptation of camera angle as a function of altitude or mission phase, further aligning with low-SWaP-C design principles that emphasize simplicity and resilience over precision hardware.

The inclusion of a cost model transforms these insights from geometric to economic relevance. The linear resolution–price relationship embedded in the optimization introduces an explicit affordability gradient, allowing cost to be traded directly against estimator performance. The resulting Pareto fronts show that MOB localization accuracy saturates below 1.0~m across all feasible designs. Beyond this threshold, additional expenditure, whether through higher-resolution sensors or larger UAV teams, yields only marginal improvements in accuracy. This defines a cost-performance boundary where further investment primarily enhances redundancy and mission persistence rather than localization accuracy.

The flattening of the Pareto surface along the MOB axis indicates that increased expenditure contributes primarily to redundancy and mission resilience rather than measurable accuracy.  Redundancy provides tangible operational benefits in contested or uncertain environments, where sensor occlusion, communication loss, or individual platform failure could otherwise jeopardize mission success. Architectures with overlapping fields of view can maintain continuous observability and mission continuity even under degraded conditions. The framework thus quantifies a fundamental design transition from accuracy-driven optimization to resilience-driven architecture design, providing engineers with a measurable criterion for when redundancy becomes the dominant performance contributor.

From a systems-engineering perspective, embedding cost as a co-optimized objective allows these trade-offs to be quantified rather than inferred qualitatively. The framework identifies the inflection point where increasing cost no longer improves accuracy but instead contributes to robustness and attritability. This capability enables mission designers to make analytically defensible decisions about how many UAVs to deploy, which camera resolution to select, and when additional spending provides diminishing returns. Because the cost model is parametric, it can be readily adapted to reflect production volume, attrition tolerance, or technology refresh cycles, extending the analysis from mission-level design to life-cycle affordability assessments.

The results of the case study further highlight the broader usefulness of the proposed mission engineering framework and how it enables designers to move beyond narrow subsystem optimization to explore the full trade space at the mission level. By embedding estimator-informed metrics such as RMSE and PCRLB into the optimization loop, the framework ensures that candidate designs are judged on their ability to meet ISR mission objectives rather than isolated technical specifications. Additionally, the joint consideration of estimator performance, mission geometry, cost, and UAV coordination within the framework exploits interactions that would have remained hidden in isolated subsystem studies. The resulting Pareto front captures the inherent trade-offs between competing objectives, providing transparency into the design space and avoiding premature commitment to a single solution. In this way, the framework does not simply yield optimal designs, but systematically reveals the structure of the trade space.

Another key aspect is the integration of the Validation Square methodology, which balances theoretical and empirical validation. Theoretical validity for this case study is provided through consistency with information-theoretic benchmarks such as the PCRLB, while empirical validity is confirmed by simulation-based estimator performance under stochastic variability. Structural realism is achieved through dynamic modeling of vehicle and sensor behaviors. Finally, a human-in-the-loop element ensures that results align with mission engineering judgment, bridging the gap between algorithmic output and operational feasibility.

From a SoS perspective, this work demonstrates how the mission engineering framework enables rapid, quantitative exploration of architectures that exploit the affordability and attritability of low-SWaP-C UAVs. By coupling optimization, simulation, and validation, the framework replaces heuristic design iteration with a repeatable, data-driven process that identifies cost-effective and resilient solutions, reducing risk and maximizing design agility early in the SoS design process.


\section{Conclusion}
This paper presented a mission engineering framework that integrates design of experiments, multi-objective optimization, and simulation into a unified process for early-phase architecture evaluation of low-SWaP-C UAVs in ISR missions. The framework systematically links design variables, estimator-informed performance metrics, and operational constraints, and cost models, enabling engineers to explore complex trade spaces that cannot be addressed by traditional platform-centric design approaches. By embedding estimator performance within the optimization loop and using Monte Carlo–derived RMSE as the primary evaluation metric, the framework ensures that design outcomes are both analytically credible and operationally relevant.

The case study on fixed-FOV UAVs conducting man-overboard localization in a GNSS-denied maritime environment demonstrates the framework’s ability to reveal the structure of the mission trade space. Results show that MOB localization accuracy rapidly saturates, achieving sub-meter performance across all configurations.  Beyond this point, higher-resolution sensors or larger UAV teams offer diminishing returns in accuracy but incur higher mission costs. The primary benefit of higher-cost configurations lies in redundancy and resilience, as overlapping sensing geometries mitigate risks from communication dropouts, occlusions, or individual platform failures. These findings emphasize that mission assurance, rather than incremental localization accuracy, becomes the dominant driver once the performance threshold is met.

The framework’s modular structure makes it readily extensible to other SoS contexts.  Vehicle dynamics, sensor models, estimator formulations, and cost relationships can be adapted or replaced without altering the overall process flow, providing a repeatable and scalable means of discovering non-obvious trade-offs among performance, cost, and resilience. The integration of the Validation Square further enhances credibility by combining theoretical, empirical, and structural validation with human-in-the-loop review, ensuring that results remain both numerically optimal and operationally feasible.

Several extensions of the framework are envisioned. First, integration with surrogate modeling and reduced-order optimization could accelerate trade space exploration while maintaining analytical fidelity. Second, embedding adaptive decision rules would allow dynamic adjustment of objectives and constraints as mission priorities evolve or cost drivers shift during operations. Finally, coupling the framework with MBSE environments would improve traceability from mission objectives to system requirements and acquisition decisions, creating a direct bridge between analytical optimization and program-level cost analysis.

In summary, the proposed mission engineering framework provides a rigorous, scalable, and extensible foundation for mission-centric SoS design. By coupling optimization, simulation, and validation in a closed loop, it enables informed, evidence-based decisions early in the design process, when flexibility is greatest and the cost of change is lowest. Beyond its technical contributions, the framework delivers managerial value by supplying transparent, data-driven trade-off analyses that link architectural choices, estimator performance, affordability, and resilience—supporting acquisition strategies and operational planning grounded in measurable system effectiveness.

\appendices
\section{Tested Camera Specifications}\label{app:Camera_specs}
Table~\ref{tab:camera_modules_digikey} provides a representative list of the camera specifications used in the DOE. 

\begin{table*}[!hbt]
\caption{Representative Camera Modules Used in DOE}
\label{tab:camera_modules_digikey}
\centering
\begin{tabular}{cccccc}
\hline
\textbf{Manufacturer} & \textbf{Part Number} & \textbf{Sensor} & \textbf{MP Class} & \textbf{Resolution (px)} & \textbf{Digi-Key Product Link} \\
\hline
OmniVision Technologies Inc. & OVM9724-RADA & -- & 1 & 1280 × 720 & \href{https://www.digikey.com/en/products/detail/omnivision-technologies-inc/OVM9724-RADA/4377196}{OmniVision OVM9724 Camera Module} \\
Image Quality Labs Inc. & IQL-OV2710/FF & OV2710 & 2 & 1920 × 1080 & \href{https://www.digikey.com/en/products/detail/image-quality-labs-inc/IQL-OV2710-FF/22022848?s=N4IgTCBcDaIJIEUAyBaA8gNTAdgIwAYB6AMWJAF0BfIA}{Image Quality Labs OV2710 Digital Camera Module} \\
Leopard Imaging Inc. & LI-OV4689-MIPI & OV4689 & 4 & 2688 × 1520 & \href{https://www.digikey.com/en/products/detail/leopard-imaging-inc/LI-OV4689-MIPI/21324072?s=N4IgTCBcDaIDIEkC0B5AagFgGwA4CcSAsggAoIgC6AvkA}{Leopard Imaging OV4689 Camera Module} \\
TechNexion & TEVI-OV5640-C & OV5640 & 5 & 2592 x 1944 & \href{https://www.digikey.com/en/products/detail/technexion/TEVI-OV5640-C/20498957}{TechNexion TEVI-OV5640 Camera Module} \\
Image Quality Labs Inc. & IQL-IMX415/FF & IMX415 & 8 & 3684 × 2176 & \href{https://www.digikey.com/en/products/detail/image-quality-labs-inc/IQL-IMX415-FF/23568700}{Sony IQL-IMX415 Camera Module} \\
OmniVision Technologies Inc. & OG09A10 & -- & 9 & 4096 x 2160 & \href{https://www.digikey.com/en/products/detail/omnivision-technologies-inc/OG09A10-C04U-001A-Z/25323276}{OmniVision OG09A10 Camera Module} \\
Camemake & CM\_USB2\_25 & IMX362 & 12 & 4000 × 3000 & \href{https://www.digikey.com/en/products/detail/camemake/CM-USB2-25/26797384?s=N4IgTCBcDaIMIFkD6BVAygITEsBWEAugL5A}{Camemake CM\_USB2\_25 Camera Module} \\
Allied Vision, Inc. & 15580 & IMX542 & 16 & 5328 × 3040 & \href{https://www.digikey.com/en/products/detail/allied-vision-inc/15580/14302502}{Alvium 1800 U-1620c Camera Module} \\
Allied Vision Inc. & 2285-14865-ND & IMX183 & 20 & 5376 x 3672 & \href{https://www.digikey.com/en/products/detail/allied-vision-inc/14865/12368301}{Alvium 1800 C-2050c Camera Module} \\
Allied Vision Inc. & 15190 & IMX540 & 25 & 5328 × 4608 & \href{https://www.digikey.com/en/products/detail/allied-vision-inc/15190/14302496}{Alvium 1800 U-2460c Camera Module} \\
\hline
\end{tabular}
\end{table*}

\section{Two-Dimensional Projections of the Pareto Front}
\label{app:ParetoProjections}
\noindent
To assist readers in visualizing the trade relationships among the three objectives shown in the main 3-D Pareto front, this appendix presents orthogonal two-dimensional projections. Each plot isolates the pairwise trade-offs between mission cost, MOB localization error, and support vessel localization error. These complementary views provide clearer insight into the structure of the Pareto-optimal solutions and their relative clustering within the trade space.

Each point represents a candidate solution from the final design population. Notably, Lower RMSE values correspond to higher estimator performance while the clustering pattern results from the optimizer converging to solutions with similar performance characteristics across different SoS architectures.

\begin{figure}[!t]
    \centering
    \includegraphics[width=0.95\linewidth]{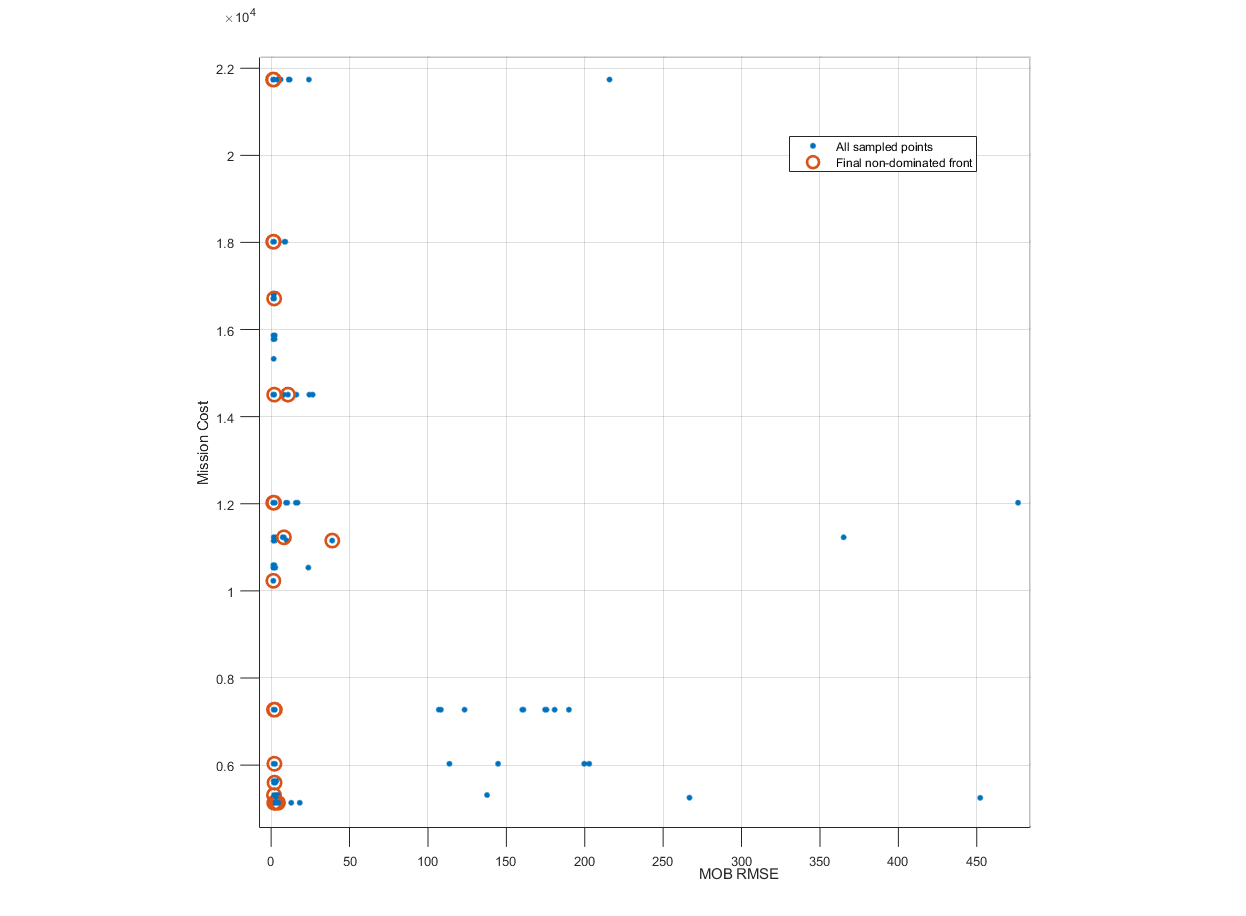}
    \caption{Pareto front projection showing the tradeoff between system cost and man-overboard MOB localization accuracy.}
    \label{fig:cost_vs_MOB_RMSE}
\end{figure}

\begin{figure}[!t]
    \centering
    \includegraphics[width=0.95\linewidth]{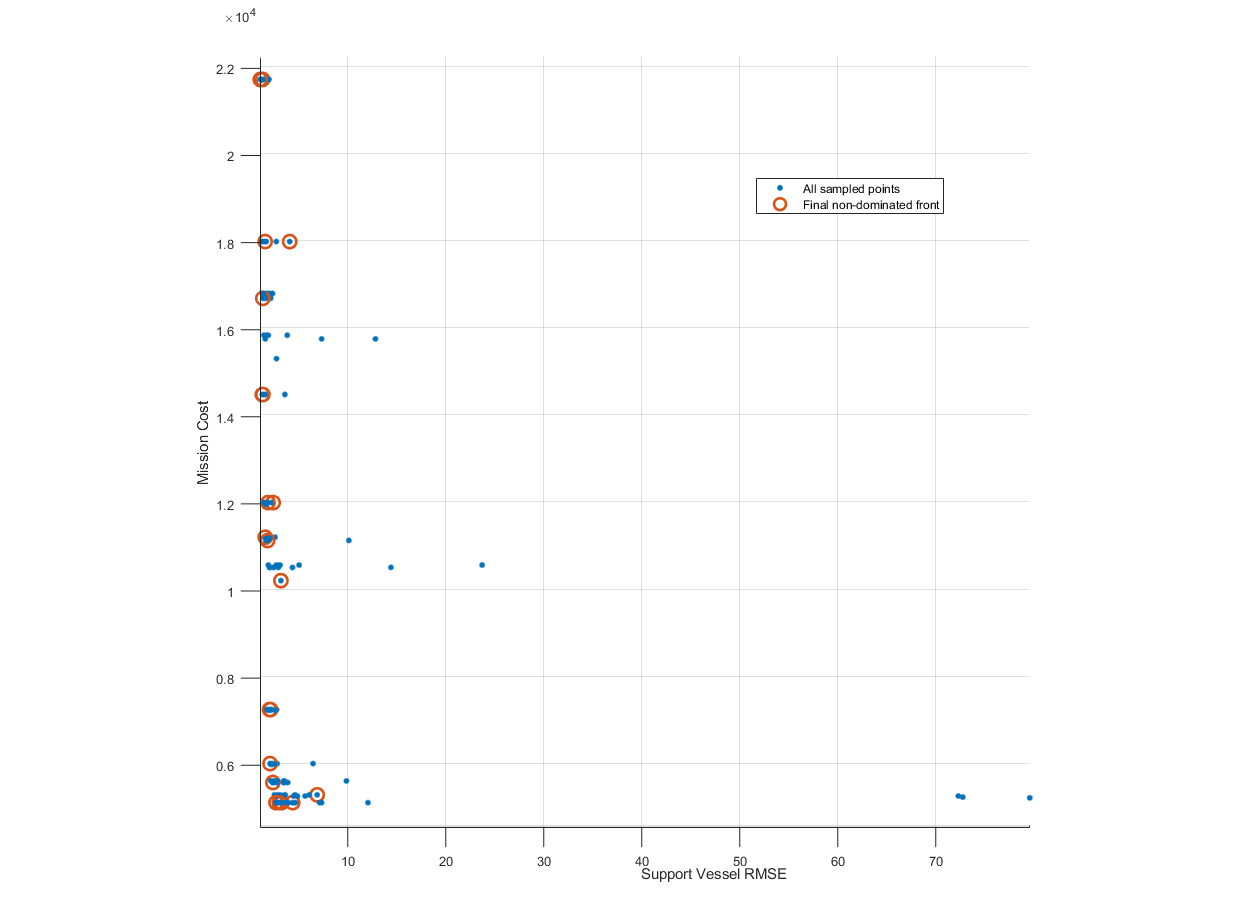}
    \caption{Pareto front projection illustrating the tradeoff between system cost and support vessel localization accuracy. }
    \label{fig:cost_vs_Support_RMSE}
\end{figure}

\begin{figure}[!t]
    \centering
    \includegraphics[width=0.95\linewidth]{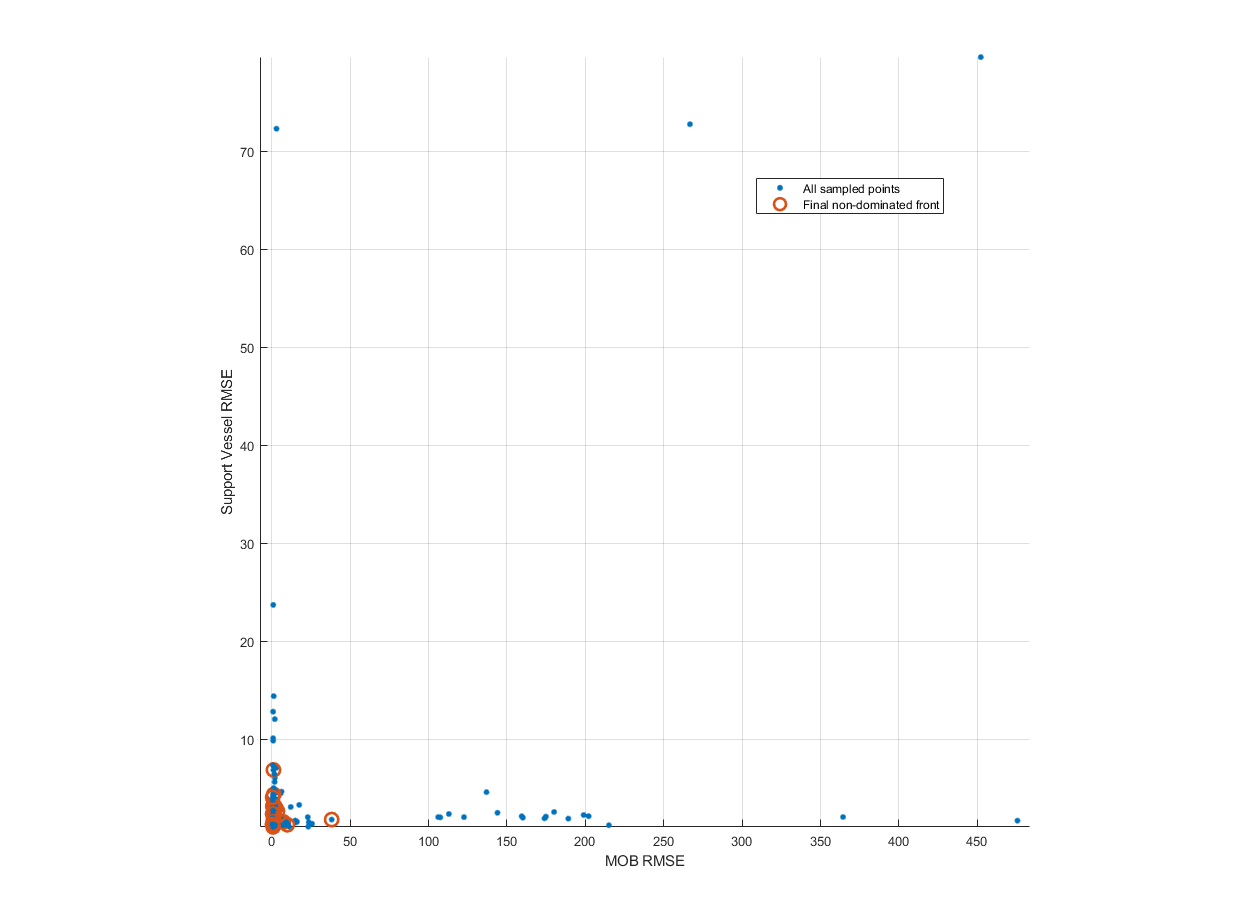}
    \caption{Projection of the Pareto front showing the coupled trade between MOB and support vessel localization errors. }
    \label{fig:MOB_vs_Support_RMSE}
\end{figure}


\section*{Data and Code Availability}
The camera module dataset derived from open-source information found on Digi-Key (obtained October 2025) and accompanying analysis scripts are available from
the corresponding author upon reasonable request. These materials include the data used in Table~\ref{tab:camera_modules_digikey}, the resolution versus cost regression data, and the MATLAB routines used for sensor trade space evaluation.

\section*{Acknowledgment}
The views presented are those of the authors and do not necessarily represent the views of the Navy, the Department of Defense, or any other agency, institution, or company. This work was authored in part by a U.S. Government employee in the scope of his/her employment. Approved for Public Release; distribution is unlimited.

\bibliographystyle{IEEEtran}
\bibliography{bibliography}

\begin{IEEEbiography}[{\includegraphics[width=1in,height=1in,clip,keepaspectratio]{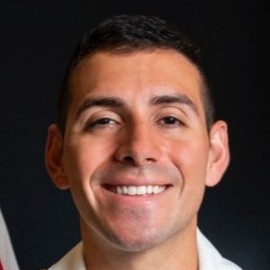}}]{Alfonso Sciacchitano}
is a doctoral candidate in the Department of Mechanical and Aerospace Engineering at the Naval Postgraduate School. His research focuses on bearings-only trajectory design, optimal estimation, and control of nonlinear and underactuated systems, with applications in autonomous and mission-critical aerospace and naval platforms. He holds a commission in the United States Navy and currently serves as an Engineering Duty Officer, with operational and technical experience supporting submarine operations. Previous assignments include multiple operational deployments, as well as instructor and staff roles at forward-deployed commands. He has been recognized for excellence in technical training and contributed to emergent response efforts in support of fleet operations. He holds a Master of Science in Electrical Engineering and a Bachelor of Science in Biomedical Engineering from Mercer University.
\end{IEEEbiography}

\begin{IEEEbiography}[{\includegraphics[width=1in,height=1.25in,clip,keepaspectratio]{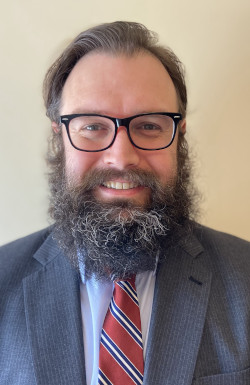}}]{Douglas L. Van Bossuyt} is an Associate Professor in the Systems Engineering Department at the Naval Postgraduate School where he is the Director of the Microgrid Innovations Research Center.  His research focuses on the nexus of failure and risk analysis, functional modeling and conceptual system design, trade-off studies and decision-making, and resilient systems.  He has published over 100 peer reviewed journal articles and conference papers, and holds several patents.  Prior to joining NPS, he was an automation engineer at a start-up company, an assistant professor at a prior institution, and a probabilistic risk assessment engineer at NuScale Power.  He was a visiting scholar at the University of Sydney Faculty of Architecture, Design, and Planning; a Space Grant Intern at the Jet Propulsion Laboratory; a DAAD-RISE Intern at the Karlsruhe Institute of Technology; and an IE3 Intern at the Centre d'\'{e}tudes maghr\'{e}bines \`{a} Tunis. He holds a Ph.D in mechanical engineering with a minor in industrial engineering; a M.S. in mechanical engineering; and a H.B.S in mechanical engineering, a H.B.A. in international studies, and a minor in business administration all from Oregon State University.
\end{IEEEbiography}
\vfill\null 



\end{document}